\documentclass[conference]{IEEEtran}

\usepackage{url}
\usepackage{array}
\usepackage{xcolor}
\usepackage{cancel}
\usepackage{siunitx}
\usepackage{fixltx2e}
\usepackage{multirow}
\usepackage{graphicx}
\usepackage{textcomp}
\usepackage{multirow}
\usepackage{stfloats}
\usepackage{algorithm}
\usepackage[T1]{fontenc}
\usepackage[noend]{algpseudocode}
\usepackage[10pt]{moresize}
\usepackage[utf8]{inputenc}
\usepackage[noadjust]{cite}
\usepackage{mathtools, cuted}
\usepackage{amsmath,amssymb,amsfonts}

\IEEEoverridecommandlockouts

\graphicspath{{./figs/}}

\newlength{\textfloatsepsave} 
\begin{document}
	
	\title{High-level Intellectual Property Obfuscation via Decoy Constants}
	
	\author{
		\IEEEauthorblockN{Levent~Aksoy\IEEEauthorrefmark{2},
			Quang-Linh~Nguyen\IEEEauthorrefmark{3},
			Felipe Almeida\IEEEauthorrefmark{2}, 
			Jaan~Raik\IEEEauthorrefmark{2},
			Marie-Lise~Flottes\IEEEauthorrefmark{3},
			Sophie~Dupuis\IEEEauthorrefmark{3} \\ and
			Samuel Pagliarini\IEEEauthorrefmark{2}}
		\IEEEauthorblockA{\IEEEauthorrefmark{2}Department of Computer Systems, Tallinn University of Technology, Tallinn, Estonia\\
			Email: \{levent.aksoy, felipe.almeida, jaan.raik, samuel.pagliarini\}@taltech.ee}
		\IEEEauthorblockA{\IEEEauthorrefmark{3}LIRMM, University of Montpellier, Montpellier, France\\
			Email: \{quang-linh.nguyen, marie-lise.flottes, sophie.dupuis\}@lirmm.fr} \vspace*{-8mm}}
	
	\maketitle
	
	\begin{abstract}
		This paper presents a high-level circuit obfuscation technique to prevent the theft of intellectual property (IP) of integrated circuits. In particular, our technique protects a class of circuits that relies on constant multiplications, such as filters and neural networks, where the constants themselves are the IP to be protected. By making use of decoy constants and a key-based scheme, a reverse engineer adversary at an untrusted foundry is rendered incapable of discerning true constants from decoy constants. The time-multiplexed constant multiplication (TMCM) block of such circuits, which realizes the multiplication of an input variable by a constant at a time, is considered as our case study for obfuscation. Furthermore, two TMCM design architectures are taken into account; an implementation using a multiplier and a multiplierless shift-adds implementation. Optimization methods are also applied to reduce the hardware complexity of these architectures. The well-known satisfiability (SAT) and automatic test pattern generation (ATPG) attacks are used to determine the vulnerability of the obfuscated designs. It is observed that the proposed technique incurs small overheads in area, power, and delay that are comparable to the hardware complexity of prominent logic locking methods. Yet, the advantage of our approach is in the insight that constants --~instead of arbitrary circuit nodes~-- become key-protected.
	\end{abstract}
	
	
	\section{Introduction}

\begin{figure*}[t]
	\centerline{\includegraphics[width=17.0cm]{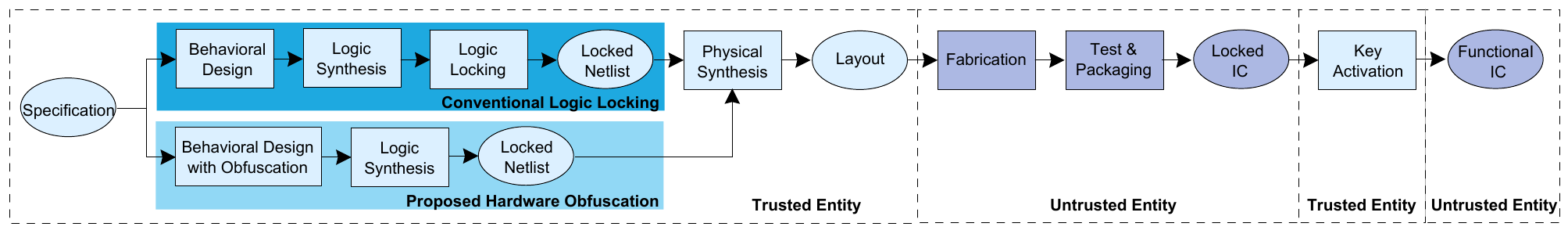}}
	\vspace*{-4mm}
	\caption{Conventional logic locking and proposed hardware obfuscation in the IC design flow (adapted from~\cite{yasin17}).}
	\label{fig:iclock}
	\vspace*{-6mm}
\end{figure*}

The involvement of multiple entities in the design and fabrication process of integrated circuits (ICs) potentially leads to security threats, such as reverse engineering, overbuilding, and insertion of malicious hardware Trojans~\cite{dsb15,torrance11,karri10}. Many efficient techniques, such as watermarking~\cite{kahng98}, IC metering~\cite{farinaz01}, IC camouflaging~\cite{raj13}, and logic locking~\cite{raj12}, have been proposed to address these issues. Among these techniques, logic locking stands out, offering a protection against a diverse array of adversaries~\cite{yasin17}. Logic locking inserts additional logic into a circuit, such as XOR/XNOR gates~\cite{roy08}, AND/OR gates~\cite{dupuis14}, or look-up tables~\cite{baumgarten10}, driven by a secret key, so that the circuit behaves as specified only when the correct key inputs are applied. The logic locking and activation of a locked circuit in the IC design flow are shown in Fig.~\ref{fig:iclock}. 

Many widely employed circuits, such as artificial neural networks (ANNs) and finite impulse response (FIR) filters, require the  multiplication of constant(s) by input variable(s). In these applications, ANN weights and filter coefficients are constants determined beforehand using sophisticated algorithms~\cite{ding18,yu10}. These constants are, therefore, an intellectual property (IP). Hence, there is a clear interest in protecting the constants since they are valuable, perhaps even more so than the circuit architecture, e.g., the number of layers in an ANN or the multiplier and accumulate block in a filter.

The hardware complexity of ANNs and filters increases as the number of neurons and filter coefficients increases, respectively, restricting their applications on design platforms with a limited number of computing resources, such as FPGAs, and on designs having a strict area requirement~\cite{nedjah09, mirzaei06}. To reduce the design area, taking into account an increase in latency, such IPs are generally implemented under a folded architecture re-using the computing resources~\cite{parhi99}. In a folded design, the time-multiplexed constant multiplication (TMCM) operation is a fundamental block that realizes the multiplication of an input variable by a single constant selected from a set of multiple constants at a time~\cite{tummeltshammer07,aksoy14_tmcm}. Since a design's layout is inevitably available to an adversary at an untrusted foundry, constants of the TMCM block are vulnerable to reverse engineering even if a logic locking method is employed.  Logic locking, despite its popularity, is not particularly well suited for hiding constants or similar design features.

Given the limitations discussed above, the main contribution of this paper is an \textbf{obfuscation technique that protects the sensitive constants from an adversary at an untrusted foundry by hiding them among decoy constants using additional logic with keyed inputs}. The proposed technique implements the obfuscation of the TMCM operation at the register-transfer level (RTL) as shown in Fig.~\ref{fig:iclock}. This enables a synthesis tool to optimize the design complexity and also promotes resource sharing, as opposed to traditional logic locking methods which are applied post synthesis at gate level. This paper considers two TMCM design architectures referred to as \mbox{{\sc tmcm-mul}} and \mbox{{\sc tmcm-sa}}. While the former utilizes multiplexors and a multiplier, the latter utilizes shifts, adders, subtractors, adders/subtractors (determined by a select input), and multiplexors under a shift-adds architecture, but no multiplier. 

The rest of this paper is organized as follows. Section~\ref{sec:background} gives the background concepts on the TMCM block and folded FIR filter design and presents the related work. The proposed TMCM obfuscation technique is described in Section~\ref{sec:architectures}. Experimental results are presented in Section~\ref{sec:results} and finally, Section~\ref{sec:conclusions} concludes the paper.



	\section{Background}
\label{sec:background}

\subsection{Time-Multiplexed Constant Multiplication}
\label{subsec:tmcm}

\begin{figure}[t]
	\centerline{\includegraphics[width=7.0cm]{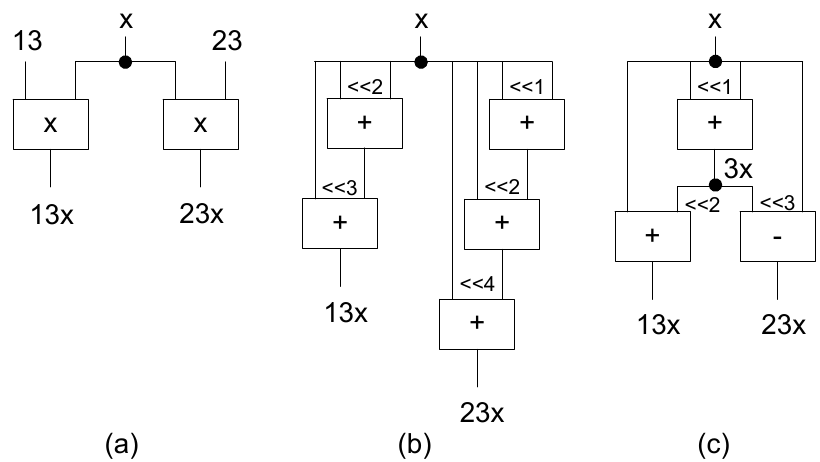}}
	\vspace*{-4mm}
	\caption{(a)~Constant multiplications of $13x$ and $23x$; their shift-adds design: (b)~DBR method~\cite{ercegovac03}; (c)~exact method~\cite{aksoy10}.}
	\label{fig:mcm}
	\vspace*{-4mm}
\end{figure}

Multiplication of constant(s) by input variable(s) is generally realized under a shift-adds architecture using only shifts, adders, and subtractors~\cite{nguyen}. Shifts by a constant value can be realized using only wires which represent no hardware cost. A straightforward way of realizing constant multiplications under a shift-adds architecture is the digit-based recoding (DBR) technique~\cite{ercegovac03}, which has two main steps: (i)~define the constants under a particular number representation, e.g., binary; (ii)~for the nonzero digits in the representation of constants, shift the input variables according to digit positions and add/subtract the shifted variables with respect to digit values. As a simple example, consider the multiplication of constants $13$ and $23$ by the input variable $x$ in the multiple constant multiplication (MCM) block shown in Fig.~\ref{fig:mcm}(a). The decompositions of constants under binary are given as follows:
\begin{align}
	13x & = (01101)_{bin}x = x \ll 3 + x \ll 2 + x \nonumber \\
	23x & = (10111)_{bin}x = x \ll 4 + x \ll 2 + x \ll 1 + x \nonumber
\end{align}
which lead to a multiplierless design with 5 operations as shown in Fig.~\ref{fig:mcm}(b). Over the years, algorithms have been proposed for minimizing the number of adders and subtractors by maximizing the sharing of partial products~\cite{aksoy14_tutorial}. For our example, the exact algorithm of~\cite{aksoy10} finds a solution with 3 operations, sharing the subexpression $3x$ as shown in Fig.~\ref{fig:mcm}(c).

\begin{figure}[t]
	\centerline{\includegraphics[width=8.7cm]{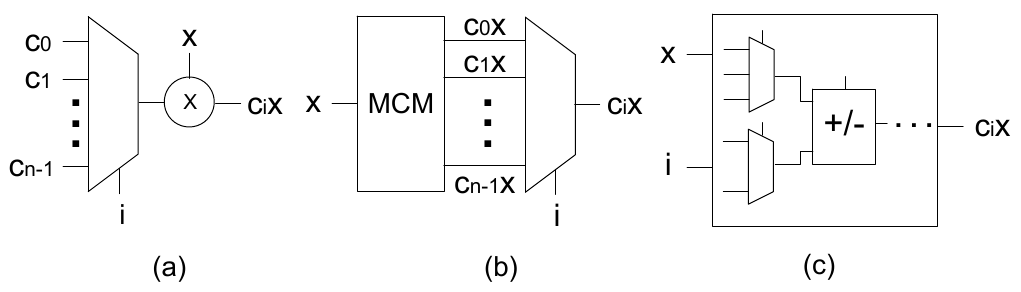}}
	\vspace*{-4mm}
	\caption{Design architectures for the implementation of the TMCM operation: (a)~\textit{mux-mul}; (b)~\textit{mcm-mux}; (c)~\textit{mux-add}.}
	\label{fig:tmcm-archs}
	\vspace*{-6mm}
\end{figure}

\begin{figure}[t]
	\centerline{\includegraphics[width=8.7cm]{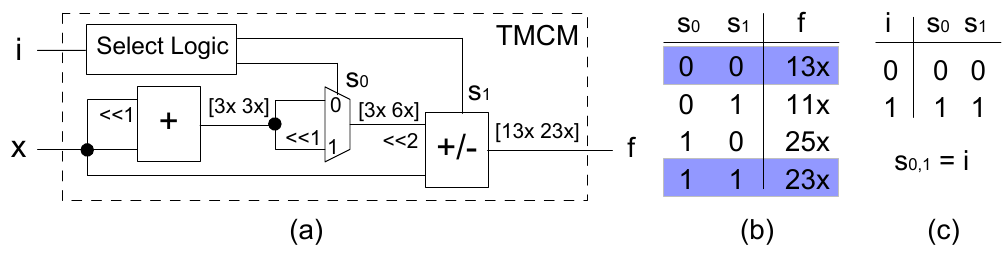}}
	\vspace*{-4mm}
	\caption{TMCM block realizing $13x$ and $23x$: (a)~solution of the algorithm~\cite{aksoy14_tmcm}; (b)~outputs based on select inputs of operations; (c)~select table.}
	\label{fig:tmcm}
	\vspace*{-6mm}
\end{figure}

On the other hand, the combinational TMCM operation can be implemented under different architectures as shown in Fig.~\ref{fig:tmcm-archs}~\cite{aksoy14_tmcm}. Given $n$ constants, the TMCM operation can be implemented using an \mbox{$n$-to-1} multiplexor and a generic multiplier, where the primary select input $i$ with \mbox{$0 \leq i \leq n-1$} determines which constant is multiplied by the input variable (cf. \mbox{\textit{mux-mul}} architecture in Fig.~\ref{fig:tmcm-archs}(a)). It can also be realized using an MCM block, which implements the multiplications of $n$ constants by the input variable, and an \mbox{$n$-to-1} multiplexor (cf. \mbox{\textit{mcm-mux}} architecture in Fig.~\ref{fig:tmcm-archs}(b)). Furthermore, it can be implemented using adders, subtractors, adders/subtractors, and multiplexors (cf. \mbox{\textit{mux-add}} architecture in Fig.~\ref{fig:tmcm-archs}(c)). Novel methods have been introduced to reduce the hardware complexity of the TMCM operation under the \mbox{\textit{mux-add}} architecture~\cite{tummeltshammer07,aksoy14_tmcm}. As a simple example, consider the TMCM operation realizing the constant multiplications of $13x$ and $23x$ at a time under the \textit{mux-add} architecture. Fig.~\ref{fig:tmcm}(a) presents the solution of the algorithm of~\cite{aksoy14_tmcm}, where the constant multiplications to be computed in time are given between square brackets in order. Note that the adder/subtractor behaves as an adder and a subtractor when its select input is 0 and 1, respectively. All possible values at the output $f$ of the TMCM operation under the select inputs of the multiplexor and adder/subtractor are given in Fig.~\ref{fig:tmcm}(b). The TMCM operation can also generate $11x$ and $25x$. To obtain the desired outputs under the primary select input $i$, the select logic is required to map $i$ to the select inputs of the multiplexor and the adder/subtractor, i.e., $s_0s_1$, as shown in Fig.~\ref{fig:tmcm}(c).

\subsection{Folded Implementation of Digital FIR Filters}

\begin{figure}[t]
	\centerline{\includegraphics[width=8.7cm]{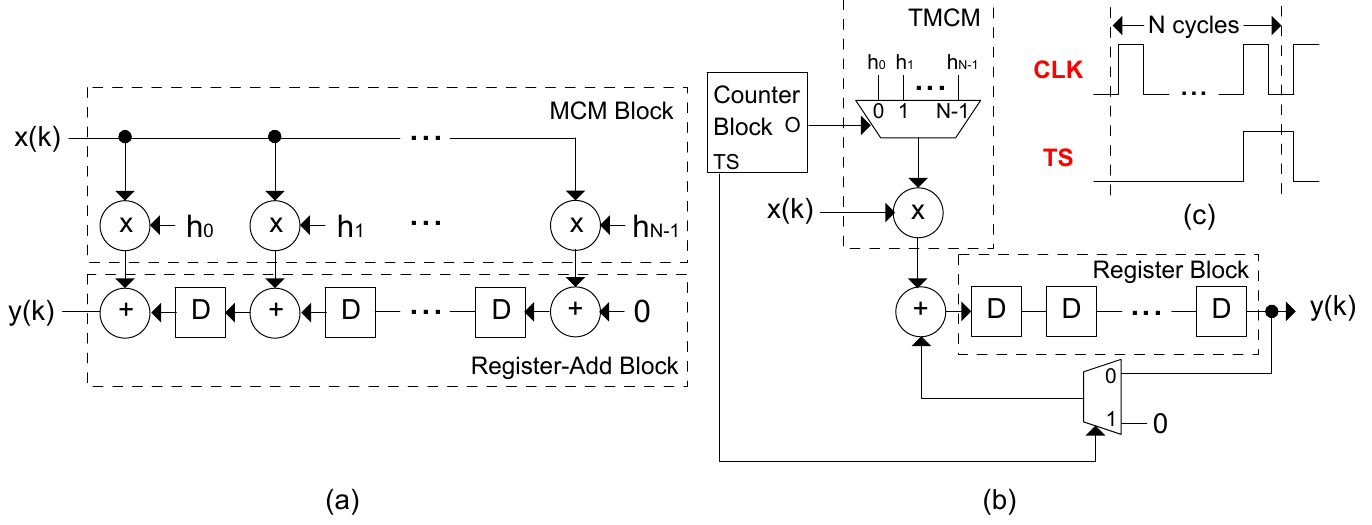}}
	\vspace*{-4mm}
	\caption{(a)~Parallel transposed form FIR filter design; (b)~its fully folded design; (c)~the timing signal $TS$.}
	\vspace*{-6mm}
	\label{fig:trans-folded}
\end{figure}

Digital filtering is frequently used in digital signal processing (DSP) applications and FIR filters are generally preferred due to their stability and linear phase property~\cite{wanhammar99}. The output of an $N$-tap FIR filter $y(k)$ is computed as $\sum_{j=0}^{N-1}h_j\cdot x(k-j)$, where $N$ is the filter length, $h_j$ is the $j^{th}$ filter coefficient, and $x(k-j)$ is the $j^{th}$ previous filter input with $0 \leq j \leq N-1$. Fig.~\ref{fig:trans-folded}(a) presents the parallel design of the transposed form FIR filter. On the other hand, Fig.~\ref{fig:trans-folded}(b) shows its folded design. The $\lceil log_2{N} \rceil$-bit counter counts from 0 to $N-1$, generating the timing signal $TS$ shown in Fig.~\ref{fig:trans-folded}(c). In this figure, $CLK$ denotes the clock signal fed to all registers which was not shown in Figs.~\ref{fig:trans-folded}(a)-(b) for the sake of clarity. The register block includes $N-1$ cascaded registers whose counterparts in the parallel design are the ones in the register-add block. Although the complexity of the filter design is reduced under the folded architecture by reusing the common operations, the filter output is obtained in $N$ clock cycles.

\subsection{Related Work}

Over the years, many defense and attack techniques have been introduced to lock logic circuits and determine the key values of the locked circuits, respectively~\cite{zamiri19}. Among the attacks against logic locking, the satisfiability (SAT) based attack~\cite{subramanyan15} is a powerful one, which was able to overcome the defenses existing at the time of its publication, such as the logic locking techniques of~\cite{roy08, dupuis14}. In recent years, a large number of SAT attack resilient logic locking methods have been introduced~\cite{yasin16,yasin17}.

Efficient hardware obfuscation techniques proposed to protect IPs have been presented in~\cite{chakraborty09, pilato08,islam20}. For the obfuscation of DSP circuits, a novel approach that uses high-level transformations, a key-based finite-state machine and a reconfigurator was introduced in~\cite{lao15}. The use of decoys in obfuscation has been utilized in~\cite{sweeney20}, but in a manner that is not related to the IP itself, as the decoys are keyed gates. To make the reverse engineering of coefficients of an FIR filter harder for an end-user, adding input and output noises was proposed in~\cite{bottegal}. To the best of our knowledge, the use of decoys to hide the target constants of an IP at RTL has not been explored before.

	\section{The Proposed Obfuscation Method}
\label{sec:architectures}

Although target constants can be stored in a tamper-proof memory as the keys in conventional logic locking, this would prevent both sharing of hardware resources and use of a multiplierless design which can lead to a significant reduction in hardware complexity when utilized as shown in Section~\ref{subsec:tmcm}. Thus, given a set of $n$ target constants $\{c_0, c_1, \ldots, c_{n-1}\}$, $m$ primary select inputs $i_0, i_1, \ldots, i_{m-1}$, where $m = \lceil log_{2}n \rceil$, and $p$ key inputs $k_0, k_1, \ldots, k_{p-1}$, the proposed obfuscation technique initially assigns decoy constants to each target constant in an iterative manner as shown in Algorithm~\ref{algo:decoyassign}. In the \textit{AssignDecoy} function of Algorithm~\ref{algo:decoyassign}, the decoy constants are preferred to have a small Hamming distance with respect to the target constant under the binary representation. This is simply because synthesis tools can also implement constant multiplications under a shift-adds architecture similar to the DBR technique~\cite{ercegovac03} and maximize the sharing of partial products among constant multiplications as shown in Fig.~\ref{fig:mcm}. The decoys are selected in between $[-2^{mbw}, 2^{mbw}-1]$, where $mbw = \lceil log_2(max\{|c_j|\}) \rceil$, \mbox{$0 \leq j \leq n-1$}, denotes the maximum bit-width of $n$ target constants. The difference on bits between the target and decoy constants starts from the least significant bit. Also, the decoys are decided to be unique to increase the obfuscation. For our example in Fig.~\ref{fig:tmcm} with $n$ is 2, given the number of key inputs $p$ is 4, for the target constant 13 $(01101)_{bin}$, the decoys 9 $(01\mathbf{0}01)_{bin}$, 12 $(0110\mathbf{0})_{bin}$, and 15 $(011\mathbf{1}1)_{bin}$ are selected and for the target constant 23 $(10111)_{bin}$, the decoys 19 $(10\mathbf{0}11)_{bin}$, 21 $(101\mathbf{0}1)_{bin}$, and 22 $(1011\mathbf{0})_{bin}$ are chosen, using a total of 6 decoys.

\begin{algorithm}[t]
	\small
	\caption{Assignment of decoys to target constants}
	\begin{algorithmic}[1]
		\Statex \textbf{Given:} $n$ target constants $\{c_0, c_1, \ldots, c_{n-1}\}$ and $p$ key inputs
		\State $noi = 0$ \Comment{Number of iterations}
		\State $nok = 0$ \Comment{Number of used keys}
		\While{$nok \ne p$}
		\State $nod = 2^{noi}$ \Comment{Number of decoys to be assigned}
		\For{$j=0$ \textbf{to}  $n-1$}
		\State AssignDecoy($c_{j}$, $nod$)
		\State $nok = nok + 1$
		\If{$nok == p$}
		\State \textbf{break}
		\EndIf
		\EndFor
		\State $noi = noi + 1$
		\EndWhile
	\end{algorithmic}
	\label{algo:decoyassign}
\end{algorithm}

\begin{figure}[t]
	\vspace*{-4mm}
	\centerline{\includegraphics[width=8.7cm]{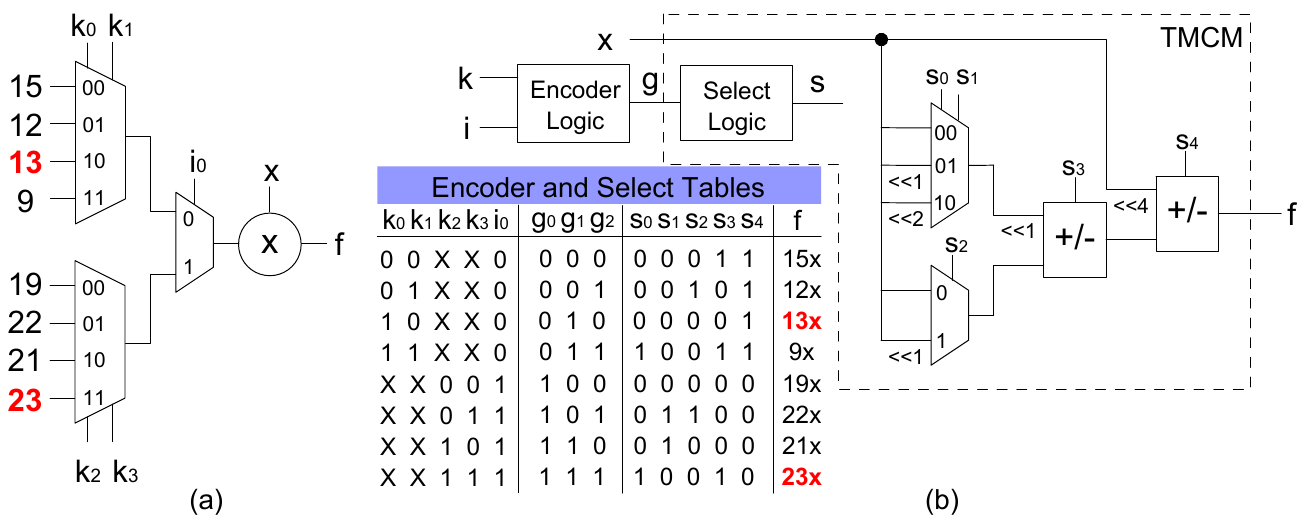}}
	\vspace*{-4mm}
	\caption{Obfuscation of the TMCM operation with target constant multiplications of $13x$ and $23x$ using decoy constants: (a)~{\sc tmcm-mul} architecture; (b)~{\sc tmcm-sa} architecture.}
	\label{fig:tmcm-reas}
	\vspace*{-6mm}
\end{figure}

Then, the obfuscated TMCM operation is implemented based on the given design architecture, i.e., {\sc tmcm-mul} or \mbox{{\sc tmcm-sa}}. The {\sc tmcm-mul} architecture is based on the \mbox{\textit{mux-mul}} architecture shown in Fig.~\ref{fig:tmcm-archs}(a). Initially, for each target constant, a multiplexor is used to select the target and its decoy constant(s) using key input(s). The locations of the target and decoy constants at the multiplexor inputs are determined randomly. The number of inputs of these multiplexors is equal to the number of decoy constants assigned to the target constant plus 1 (denoting the target constant). Then, another multiplexor is used to realize the constant multiplication in the TMCM block in the given order using the primary select input $i$. The number of inputs of this multiplexor is equal to $n$. Finally, a multiplier is used to realize the constant multiplication. Its size is equal to $mbw + ibw$, where $ibw$ denotes the bit-width of the input variable $x$. For our example in Fig.~\ref{fig:tmcm}, the realization of the obfuscated TMCM block is illustrated in Fig.~\ref{fig:tmcm-reas}(a). The constant multiplications $13x$ and $23x$ are computed when \mbox{$k_0k_1k_2k_3i_0 = 10XX0$} and \mbox{$k_0k_1k_2k_3i_0 = XX111$}, respectively, where $X$ is the don't care. Note that a wrong key leads to at least one decoy multiplication at a time, e.g., if $k_0k_1k_2k_3 = 1000$, the TMCM block generates $13x$ and $19x$ when $i_0$ is 0 and 1, respectively. Hence, in a folded design such as FIR filter in Fig.~\ref{fig:trans-folded}(b), whose  TMCM block is obfuscated using the proposed method, a wrong key always generates a wrong output.

The \textit{\sc tmcm-sa} architecture is based on the \textit{mux-add} architecture shown in Fig.~\ref{fig:tmcm-archs}(c). Initially, given the order of target and decoy constant multiplications, an encoder logic, which maps the key inputs $k$ and primary select $i$ to the select inputs of the TMCM operation, i.e., $g$, is generated. Note that the encoder logic has $m+p$ inputs and $\lceil log_{2}(n+r) \rceil$ outputs, where $r$ denotes the total number of decoy constants. Then, the constant multiplications in the same order are given to the algorithm of~\cite{aksoy14_tmcm} and an optimized implementation is found under the shift-adds architecture. For our simple example, given the order of target and decoy constant multiplications as in Fig.~\ref{fig:tmcm-reas}(a), i.e., $[15x~12x~13x~9x~19x~22x~21x~23x]$, the realization of the obfuscated TMCM operation is shown in Fig.~\ref{fig:tmcm-reas}(b) including the encoder and select tables. Note that the select logic, which is implemented by the algorithm of~\cite{aksoy14_tmcm} to realize the constant multiplications in a given order, maps the select inputs of the TMCM operation, i.e., $g$, to the select inputs of adders/subtractors and multiplexors in the TMCM design, i.e., $s_0s_1s_2s_3s_4$. 

The TMCM operation can also be obfuscated based on the \mbox{\textit{mcm-mux}} architecture shown in Fig.~\ref{fig:tmcm-archs}(b). In our obfuscated implementation, the MCM block realizes the target and decoy constant multiplications. For each target constant with decoys, a multiplexor is used to select the constant multiplication using the key inputs. Finally, the multiplexor with the primary select input $i$, which generates the output, is used. However, in this case, it was observed that the key inputs are so vulnerable to the logic locking attacks since they can be observed easily at the output. The key inputs in the \mbox{\sc tmcm-mul} and \mbox{\sc tmcm-sa} architectures are behind many logic operators, which is preferred from an obfuscation point of view.

To automate the design and verification process, a computer-aided design (CAD) tool was developed for the obfuscation of the TMCM operation. The CAD tool takes the target constants, the number of key inputs, and the design architecture as inputs, and generates the description of the obfuscated TMCM design in Verilog, the testbench for verification, and synthesis and simulation scripts. Note that the designs under both architectures are described in a behavioral fashion. While targets and decoys are expressed as constants in the RTL code under the {\sc tmcm-mul} architecture, the TMCM block, which realizes the multiplication of the input variable by a target or decoy constant at a time, is generated by the method of~\cite{aksoy14_tmcm} under the {\sc tmcm-sa} architecture. 

	\section{Experimental Results}
\label{sec:results}

As an experiment set, three FIR filters, selected from the FIR filter benchmark suite~\cite{firsuite}, were used. Their specifications are given in Table~\ref{tab:firs}, where $n$ and $m$ are the number of filter coefficients and primary select inputs, repectively, and $mbw$ is the maximum bit-width of filter coefficients. Note that \textit{\#in} and \textit{\#out} stand respectively for the number of inputs and outputs of the TMCM design computed when the bit-width of the filter input $ibw$ is 32. 

\begin{table}[t]
	\centering
	\caption{Specifications of FIR filters and TMCM blocks.} 
	\vspace*{-3mm}
	\begin{tabular}{|cc||ccc||cc|}
		\hline
		\multicolumn{2}{|c||}{Filter} & \multicolumn{3}{c||}{Coefficient Details} &  \multicolumn{2}{c|}{TMCM Details} \\
		\hline
		Index & Name & n & m & mbw & \#in & \#out\\
		\hline \hline
		1 & \multicolumn{1}{l||}{Johansson08\_30} & 30  & 5 & 10 & 37 & 42\\
		2 & \multicolumn{1}{l||}{Shi11\_S2}       & 60  & 6 & 10 & 38 & 42\\
		3 & \multicolumn{1}{l||}{Maskell07\_A108} & 108 & 7 & 9  & 39 & 41\\
		\hline
	\end{tabular}
	\label{tab:firs}
	\vspace*{-6mm}
\end{table}

Conventional logic locking was applied to the combinational TMCM blocks of FIR filters under the \mbox{\textit{mux-mul}} and \mbox{\textit{mux-add}} architectures, while the proposed technique was used to obfuscate these combinational TMCM blocks under the \mbox{\sc tmcm-mul} and \mbox{\sc tmcm-sa} architectures. Logic synthesis was performed by Cadence Genus using a commercial 65nm cell library. The functionality of designs was verified on 10,000 randomly generated input signals in simulation, from which the switching activity information is collected and utilized by the synthesis tool to compute power dissipation. The SAT and automatic test pattern generation (ATPG) based attacks developed in~\cite{subramanyan15} were used to check the resiliency of locked and obfuscated TMCM designs after they were synthesized into a gate-level netlist. Note that the ATPG based attack of~\cite{subramanyan15} initially uses ATPG methods to determine key inputs and then, the SAT based attack to find the rest of key inputs undetermined by ATPG methods. As a common practice, a time limit of 1 day was given to the attacks which were run on a computing cluster including Intel Xeon processing units at 2.4GHz with 40 cores and 96GB memory. 

Tables~\ref{tab:locked} and~\ref{tab:obfuscated} present the gate-level design results of locked and obfuscated combinational TMCM blocks, respectively, when $ibw$ is 32. The number of key inputs $p$ is 32, 64, and 128 for the FIR filter \textit{Johansson08\_30}, \textit{Shi11\_S2}, and \textit{Maskell07\_A108}, respectively. In these tables, \textit{area}, \textit{delay}, and \textit{power} stand for the total area in $\mu m^2$, delay in the critical path in $ps$, and total power dissipation in $\mu W$, respectively. Also, {\sc Asat} and {\sc Aatpg} denote the SAT and ATPG based attacks, respectively, where \textit{time} is their run-time in seconds. In Table~\ref{tab:locked}, \textit{LLT} denotes a logic locking technique, where {\sc rand}, {\sc iolts}, {\sc sar}, and {\sc sfll}\footnote{The $h$ parameter of the SFLL-HD technique, which is used to adjust the tradeoff between SAT attack resiliency and output corruption, was set to $p/4$ in our experiments.} are the methods of~\cite{roy08},~\cite{dupuis14},~\cite{yasin16}, and~\cite{yasin17}, respectively.

\begin{table}[t]
	\centering
	\scriptsize
	\caption{Results of TMCM blocks of FIR filters locked by prominent techniques.}
	\vspace*{-3mm}
	\begin{tabular}{|@{\hskip2pt}c@{\hskip2pt}|l|l||c|c|c||c|c|}
		\hline
		Filter & \multirow{2}{*}{Architecture} & \multirow{2}{*}{LLT} & \multicolumn{3}{c||}{Synthesis Results} & \multicolumn{1}{c|}{\sc Asat}  & \multicolumn{1}{c|}{\sc Aatpg} \\
		\cline{4-8}
		Index & & & area & delay & power & time & time \\
		\hline \hline
		\multirow{8}{*}{1}  & \multirow{4}{*}{\textit{mux-mul}} & {\sc rand}  & 2775 & 5558 & 1147 & 49743 & 22\\
		&                                   & {\sc iolts} & 2802 & 4355 & 1229 & >1day & >1day\\
		&                                   & {\sc sar}   & 2927 & 4436 & 1225 & >1day & >1day\\
		&                                   & {\sc sfll}  & 4020 & 4473 & 2391 & 301   & 100\\
		\cline{2-8}                                                    
		& \multirow{4}{*}{\textit{mux-add}} & {\sc rand}  & 3071 & 6336 & 1393 & 65002 & 1377\\
		&                                   & {\sc iolts} & 2894 & 5234 & 1386 & >1day & >1day\\
		&                                   & {\sc sar}   & 2999 & 5122 & 1390 & >1day & >1day\\
		&                                   & {\sc sfll}  & 4092 & 5202 & 2696 & 7215 & 367\\
		\hline \hline       
		\multirow{8}{*}{2}  & \multirow{4}{*}{\textit{mux-mul}} & {\sc rand}  & 3104 & 5588 & 1334 & >1day & 18\\
		&                                   & {\sc iolts} & 3035 & 4558 & 1279 & >1day & >1day\\
		&                                   & {\sc sar}   & 3168 & 4603 & 1236 & >1day & >1day\\
		&                                   & {\sc sfll}  & 5648 & 6290 & 4998 & >1day & >1day\\
		\cline{2-8}                                                   
		& \multirow{4}{*}{\textit{mux-add}} & {\sc rand}  & 3570 & 6661 & 1536 & >1day & >1day\\
		&                                   & {\sc iolts} & 3260 & 5939 & 1400 & >1day & >1day\\
		&                                   & {\sc sar}   & 3395 & 5730 & 1397 & >1day & >1day\\
		&                                   & {\sc sfll}  & 5913 & 6136 & 5021 & >1day & >1day\\
		\hline \hline                                                 
		\multirow{8}{*}{3}  & \multirow{4}{*}{\textit{mux-mul}} & {\sc rand}  & 3020 & 6268  & 1186  & 50758 & 33327\\
		&                                   & {\sc iolts} & 2759 & 4765  & 1112  & 21104 & 23423\\
		&                                   & {\sc sar}   & 3070 & 4945  & 1121  & >1day & >1day\\
		&                                   & {\sc sfll}  & 9313 & 11629 & 17585 & >1day & >1day\\
		\cline{2-8}                                                   
		& \multirow{4}{*}{\textit{mux-add}} & {\sc rand}  & 3913  & 6705  & 1126  & 10555 & 7254 \\
		&                                   & {\sc iolts} & 3462  & 5189  & 1099  & 19235 & 20885\\
		&                                   & {\sc sar}   & 3703  & 5149  & 1070  & >1day & >1day\\
		&                                   & {\sc sfll}  & 10000 & 11909 & 17487 & >1day & >1day\\
		\hline
	\end{tabular}
	\label{tab:locked}
	\vspace*{-4mm}
\end{table}

\begin{table}[t]
	\centering
	\scriptsize
	\caption{Results of TMCM blocks of FIR filters obfuscated by the proposed technique.}
	\vspace*{-3mm}
	\begin{tabular}{|c|l||c|c|c||c|c|}
		\hline
		Filter & \multirow{2}{*}{Architecture} & \multicolumn{3}{c||}{Synthesis Results} & \multicolumn{1}{c|}{\sc Asat}  & \multicolumn{1}{c|}{\sc Aatpg} \\
		\cline{3-7}
		Index & & area & delay & power & time & time \\
		\hline \hline
		\multirow{2}{*}{1} & {\sc tmcm-mul} & 2749 & 5341 & 1170 & >1day & >1day\\
		& {\sc tmcm-sa}  & 3445 & 6651 & 2024 & >1day & >1day\\
		\hline \hline
		\multirow{2}{*}{2} & {\sc tmcm-mul} & 4362 & 5810 & 3546 & >1day & >1day\\
		& {\sc tmcm-sa}  & 4318 & 7139 & 2169 & >1day & >1day\\
		\hline \hline
		\multirow{2}{*}{3} & {\sc tmcm-mul} & 4595 & 5636 & 3522 & >1day & >1day\\
		& {\sc tmcm-sa}  & 4155 & 6256 & 1895 & >1day & >1day\\
		\hline
	\end{tabular}
	\label{tab:obfuscated}
	\vspace*{-6mm}
\end{table}

\begin{figure}[t]
	\vspace*{-4mm}
	\centerline{\includegraphics[width=8.3cm]{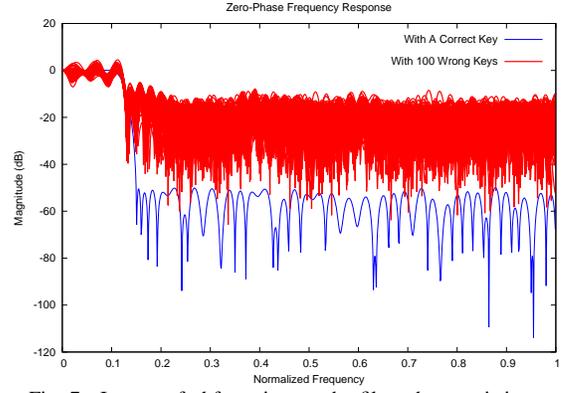}}
	\vspace*{-10mm}
	\caption{Impact of obfuscation on the filter characteristics.}
	\label{fig:zpfr}
	\vspace*{-6mm}
\end{figure}

Observe from Table~\ref{tab:locked} that while the TMCM designs locked by {\sc rand}, {\sc iolts} and {\sc sfll} techniques are vulnerable to the SAT and ATPG based attacks, the {\sc sfll} technique leads to locked designs which have larger hardware complexity than those of any other techniques used in this paper. Besides, the {\sc sar} method generates locked designs comparable to those generated by the {\sc iolts} and {\sc rand} methods in terms of hardware complexity, but more resilient to the attacks. On the other hand, observe from Table~\ref{tab:obfuscated} that the proposed obfuscation technique generates locked designs that the SAT and ATPG based attacks find hard to decrypt. Moreover, the designs obfuscated by the proposed technique have less hardware complexity than those locked by the {\sc sfll} technique. When compared to the designs locked by the {\sc sar} technique\footnote{Designs locked by both {\sc sar} and the proposed obfuscation methods could not be decrypted by the SAT and ATPG based logic locking attacks in a given time-limit.} under the \mbox{\textit{mux-mul}} architecture, the area, delay, and power dissipation can increase up to 49\%, 26\%, and 214\% in the obfuscated TMCM designs under the \mbox{{\sc tmcm-mul}} architecture. However, the obfuscation technique can find a TMCM design with less area on the filter \textit{Johansson08\_30}. With respect to the synthesis results of designs locked by the {\sc sar} technique under the \mbox{\textit{mux-add}} architecture, the area, delay, and power dissipation can increase up to 27\%, 29\%, and 77\% in the obfuscated designs under the \mbox{{\sc tmcm-sa}} architecture. However, we note that the main advantage of the proposed obfuscation method is to protect the target constants from reverse engineering, which cannot be guaranteed by a traditional logic locking technique. Also, {\sc sar} has an extremely low impact on circuit functionality, meaning that for each input pattern, there is only one key value that leads to corruption at outputs, whereas, our method has a high impact as shown in Fig.~\ref{fig:zpfr}. Furthermore, observe from Table~\ref{tab:obfuscated} that the \mbox{\sc tmcm-sa} architecture can lead to designs with less area and power consumption when compared to the \mbox{\sc tmcm-mul} architecture, e.g., the FIR filter \textit{Shi11\_S2} and \textit{Maskell07\_A108}. The delay of obfuscated designs under the \mbox{\sc tmcm-sa} architecture is generally larger than that of designs under the \mbox{\sc tmcm-mul} architecture because of a large number of operations in series in the TMCM blocks obtained by the algorithm of~\cite{aksoy14_tmcm}. 

To explore the impact of the number of key inputs on the hardware complexity and resiliency to the attacks of the obfuscated designs, the TMCM block of the FIR filter \textit{Johansson08\_30} is implemented with a $p$ value in between 16 and 80, increased in a step of 16. Table~\ref{tab:nok} shows the synthesis results of TMCM designs and solutions of the attacks. 

\begin{table}[t]
	\centering
	\scriptsize
	\caption{Impact of the number of key inputs in the proposed obfuscation technique.}
	\vspace*{-3mm}
	\begin{tabular}{|c|c||c|c|c||c|c|}
		\hline
		\multirow{2}{*}{$p$} & \multirow{2}{*}{Architecture} & \multicolumn{3}{c||}{Synthesis Results} & \multicolumn{1}{c|}{\sc Asat}  & \multicolumn{1}{c|}{\sc Aatpg} \\
		\cline{3-7}
		& & area & delay & power & time & time \\
		\hline \hline
		\multirow{2}{*}{16} & \multicolumn{1}{l||}{\sc tmcm-mul} & 2688 & 5390 & 1084 & 773   & 26 \\
		& \multicolumn{1}{l||}{\sc tmcm-sa}  & 3398 & 5442 & 1647 & >1day & 34 \\
		\hline
		\multirow{2}{*}{32} & \multicolumn{1}{l||}{\sc tmcm-mul} & 2749 & 5341 & 1170 & >1day & >1day \\
		& \multicolumn{1}{l||}{\sc tmcm-sa}  & 3445 & 6651 & 2024 & >1day & >1day \\
		\hline
		\multirow{2}{*}{48} & \multicolumn{1}{l||}{\sc tmcm-mul} & 2862 & 5696 & 1282 & >1day & >1day \\
		& \multicolumn{1}{l||}{\sc tmcm-sa}  & 3611 & 6586 & 1720 & >1day & >1day \\
		\hline
		\multirow{2}{*}{64} & \multicolumn{1}{l||}{\sc tmcm-mul} & 4420 & 5762 & 3132 & >1day & >1day \\
		& \multicolumn{1}{l||}{\sc tmcm-sa}  & 3697 & 7224 & 1874 & >1day & >1day \\
		\hline
		\multirow{2}{*}{80} & \multicolumn{1}{l||}{\sc tmcm-mul} & 4631 & 5656 & 3241 & >1day & >1day \\
		& \multicolumn{1}{l||}{\sc tmcm-sa}  & 4515 & 7618 & 2990 & >1day & >1day \\
		\hline
	\end{tabular}
	\label{tab:nok}
	\vspace*{-4mm}
\end{table}

Observe from Table~\ref{tab:nok} that as $p$ increases, the hardware complexity of the obfuscated designs is increased. However, the increase ratio in area and power dissipation of the designs under the \mbox{\sc tmcm-mul} architecture is larger than that of the designs under the \mbox{\sc tmcm-sa} architecture, such that the area and power dissipation of the design under the \mbox{\sc tmcm-sa} architecture become smaller than those of designs under the \mbox{\sc tmcm-mul} architecture as $p$ increases. This is because as $p$ increases, the size of multiplexors increases under the \mbox{\sc tmcm-mul} architecture and the complexity of the TMCM design under the \mbox{\sc tmcm-sa} architecture increases slightly due to the optimization algorithm of~\cite{aksoy14_tmcm}. Also, as $p$ decreases, the obfuscated TMCM design becomes less resilient to attacks. 



To find the impact of the filter input bit-width on the hardware complexity and resiliency to the attacks of the obfuscated designs, the TMCM block of the FIR filter \textit{Shi11\_S2} is designed when $p$ is 64 and $ibw$ is in between 16 and 32, increased in a step of 4. Table~\ref{tab:iw} shows the synthesis results of the obfuscated TMCM designs and solutions of the attacks. 

\begin{table}[t]
	\centering
	\scriptsize
	\caption{Impact of the bit-width of filter input in the proposed obfuscation technique.}
	\vspace*{-3mm}
	\begin{tabular}{|c|c||c|c|c||c|c|}
		\hline
		\multirow{2}{*}{$ibw$} & \multirow{2}{*}{Architecture} & \multicolumn{3}{c||}{Synthesis Results} & \multicolumn{1}{c|}{\sc Asat}  & \multicolumn{1}{c|}{\sc Aatpg} \\
		\cline{3-7}
		& & area & delay & power & time & time \\
		\hline \hline
		\multirow{2}{*}{16} & \multicolumn{1}{l||}{\sc tmcm-mul} & 2145 & 3611 & 980  & 177   & 17902 \\
		& \multicolumn{1}{l||}{\sc tmcm-sa}  & 2741 & 5291 & 1269 & 409   & 1181  \\
		\hline
		\multirow{2}{*}{20} & \multicolumn{1}{l||}{\sc tmcm-mul} & 2589 & 4670 & 1466 & 1369  & >1day \\
		& \multicolumn{1}{l||}{\sc tmcm-sa}  & 3127 & 5691 & 1555 & 2682  & 3626  \\
		\hline
		\multirow{2}{*}{24} & \multicolumn{1}{l||}{\sc tmcm-mul} & 3144 & 4603 & 1970 & 28003 & >1day \\
		& \multicolumn{1}{l||}{\sc tmcm-sa}  & 3507 & 6159 & 1802 & 13148 & 7848  \\
		\hline
		\multirow{2}{*}{28} & \multicolumn{1}{l||}{\sc tmcm-mul} & 3836 & 5000 & 2700 & >1day & >1day \\
		& \multicolumn{1}{l||}{\sc tmcm-sa}  & 3855 & 6644 & 2087 & 27221 & 31877 \\
		\hline
		\multirow{2}{*}{32} & \multicolumn{1}{l||}{\sc tmcm-mul} & 4362 & 5810 & 3546 & >1day & >1day \\
		& \multicolumn{1}{l||}{\sc tmcm-sa}  & 4318 & 7139 & 2169 & >1day & >1day \\
		\hline
	\end{tabular}
	\label{tab:iw}
	\vspace*{-6mm}
\end{table}

Observe from Table~\ref{tab:iw} that as $ibw$ increases, the hardware complexity of the TMCM designs increases since the size of operators increases. Note that the size of a multipler under the \mbox{\sc tmcm-mul} architecture has a larger impact on the hardware complexity when compared to the size of adders, subtractors, and adders/subtractors under the \mbox{\sc tmcm-sa} architecture since area and power dissipation of designs under the \mbox{\sc tmcm-mul} architecture become larger than those of designs under the \mbox{\sc tmcm-sa} architecture as $ibw$ increases. Observe that the resiliency of obfuscated designs to the attacks also increases as $ibw$ increases. This is mainly because the search space of the problem of finding key inputs increases as $ibw$ increases.

To explore the impact of the decoy selection method (DSM) on the hardware complexity and resiliency to the attacks of the obfuscated designs, the TMCM blocks of FIR filters are also implemented when decoys are chosen randomly, respecting $mbw$. In these designs, $ibw$ is 32 and $p$ is 32, 64, and 128 for the FIR filter \textit{Johansson08\_30}, \textit{Shi11\_S2}, and \textit{Maskell07\_A108}, respectively. Table~\ref{tab:dms} presents the synthesis results of the obfuscated TMCM designs obtained using different DSMs and the solutions of the attacks. 

\begin{table}[t]
	\centering
	\scriptsize
	\caption{Impact of the decoy selection method in the proposed obfuscation technique.}
	\vspace*{-3mm}
	\begin{tabular}{|@{\hskip2pt}c@{\hskip2pt}|l|c||c|c|c||c|c|}
		\hline
		Filter & \multirow{2}{*}{Architecture} & \multirow{2}{*}{DSM} & \multicolumn{3}{c||}{Synthesis Results} & \multicolumn{1}{c|}{\sc Asat}  & \multicolumn{1}{c|}{\sc Aatpg} \\
		\cline{4-8}
		Index & & & area & delay & power & time & time \\
		\hline \hline
		\multirow{4}{*}{1} & \multirow{2}{*}{\sc tmcm-mul} & \multicolumn{1}{l||}{random} & 4382 & 5388 & 3115 & >1day & 3580 \\
		&                               & \multicolumn{1}{l||}{proposed}  & 2749 & 5341 & 1170 & >1day & >1day\\
		\cline{2-8}
		& \multirow{2}{*}{\sc tmcm-sa}  & \multicolumn{1}{l||}{random} & 3489 & 5360 & 1767 & >1day & >1day\\
		&                               & \multicolumn{1}{l||}{proposed}  & 3445 & 6651 & 2024 & >1day & >1day\\
		\hline \hline
		\multirow{4}{*}{2} & \multirow{2}{*}{\sc tmcm-mul} & \multicolumn{1}{l||}{random} & 4740 & 5756 & 3784 & >1day & >1day\\
		&                               & \multicolumn{1}{l||}{proposed}  & 4362 & 5810 & 3546 & >1day & >1day\\
		\cline{2-8}
		& \multirow{2}{*}{\sc tmcm-sa}  & \multicolumn{1}{l||}{random} & 4475 & 6208 & 1800 & 26299 & 42029\\
		&                               & \multicolumn{1}{l||}{proposed}  & 4318 & 7139 & 2169 & >1day & >1day\\
		\hline \hline
		\multirow{4}{*}{3} & \multirow{2}{*}{\sc tmcm-mul} & \multicolumn{1}{l||}{random} & 4833 & 5756 & 3442 & >1day & >1day\\
		&                               & \multicolumn{1}{l||}{proposed}  & 4595 & 5636 & 3522 & >1day & >1day\\
		\cline{2-8}
		& \multirow{2}{*}{\sc tmcm-sa}  & \multicolumn{1}{l||}{random} & 4073 & 7529 & 1644 & 42414 & >1day\\
		&                               & \multicolumn{1}{l||}{proposed}  & 4155 & 6256 & 1895 & >1day & >1day\\
		\hline
	\end{tabular}
	\label{tab:dms}
	\vspace*{-4mm}
\end{table}

\begin{table}[t]
	\centering
	\scriptsize
	\caption{Impact of obfuscation in the folded FIR filter design.}
	\vspace*{-3mm}
	\begin{tabular}{|c|l|c||c|c|c|}
		\hline
		Filter & \multirow{2}{*}{Technique} & \multirow{2}{*}{Architecture} & \multicolumn{3}{c|}{Synthesis Results} \\
		\cline{4-6}
		Index & & & area & delay & power \\
		\hline \hline
		\multirow{4}{*}{1} & \multirow{2}{*}{Original}   & \multicolumn{1}{l||}{\textit{mux-mul}} & 14082 & 6362 & 1386 \\
		&                             & \multicolumn{1}{l||}{\textit{mux-add}} & 14396 & 6797 & 1592 \\
		\cline{2-6}
		& \multirow{2}{*}{Obfuscated} & \multicolumn{1}{l||}{\sc tmcm-mul}     & 14171 & 6101 & 1386 \\
		&                             & \multicolumn{1}{l||}{\sc tmcm-sa}      & 14973 & 7548 & 1969 \\
		\hline \hline
		\multirow{4}{*}{2} & \multirow{2}{*}{Original}   & \multicolumn{1}{l||}{\textit{mux-mul}} & 25007 & 6045 & 2011 \\
		&                             & \multicolumn{1}{l||}{\textit{mux-add}} & 25485 & 6789 & 2242 \\
		\cline{2-6}
		& \multirow{2}{*}{Obfuscated} & \multicolumn{1}{l||}{\sc tmcm-mul}     & 26486 & 6298 & 2382 \\
		&                             & \multicolumn{1}{l||}{\sc tmcm-sa}      & 26813 & 8172 & 2722 \\
		\hline \hline
		\multirow{4}{*}{3} & \multirow{2}{*}{Original}   & \multicolumn{1}{l||}{\textit{mux-mul}} & 41391 & 6260 & 3200 \\
		&                             & \multicolumn{1}{l||}{\textit{mux-add}} & 42238 & 6652 & 3228 \\
		\cline{2-6}
		& \multirow{2}{*}{Obfuscated} & \multicolumn{1}{l||}{\sc tmcm-mul}     & 43379 & 6568 & 3630 \\
		&                             & \multicolumn{1}{l||}{\sc tmcm-sa}      & 43738 & 8422 & 4083 \\
		\hline
	\end{tabular}
	\label{tab:firsynth}
	\vspace*{-6mm}
\end{table}

Observe from Table~\ref{tab:dms} that because the proposed DSM favors unique decoy constants with a small Hamming distance value with respect to the associated target constant, it generally leads to obfuscated TMCM designs with a smaller area when compared to the random decoy selection under both design architectures, except the FIR filter \textit{Maskell07\_A108} under the {\sc tmcm-sa} architecture. Random decoy selection can also create vulnerabilities on the key inputs which is anticipated due to the use of common decoys for different target constants. 

To investigate the impact of obfuscation of the TMCM block in filter design, these FIR filters are implemented as shown in Fig.~\ref{fig:trans-folded}(b) when their TMCM blocks are realized under the \mbox{\textit{mux-mul}} and \mbox{\textit{mux-add}} architectures and these designs are compared with filters including the obfuscated TMCM blocks generated under the \mbox{\sc tmcm-mul} and \mbox{\sc tmcm-sa} architectures. Table~\ref{tab:firsynth} shows the synthesis results obtained when $ibw$ is 32 and $p$ is 32, 64, and 128 for the FIR filter \textit{Johansson08\_30}, \textit{Shi11\_S2}, and \textit{Maskell07\_A108}, respectively.

Observe from Table~\ref{tab:firsynth} that the hardware obfuscation on the TMCM block of an FIR filter under the \mbox{{\sc tmcm-mul}} (\mbox{\sc tmcm-sa}) architecture can respectively increase the area, delay, and power dissipation up to 5.5\% (4.9\%), 4.6\% (21.0\%), and 15.5\% (20.9\%) when compared to the original FIR filters under the \textit{mux-mul} (\textit{mux-add}) architecture. Note that the TMCM block has less impact on the area of a folded FIR filter with respect to the registers. Hence, an FIR filter can be obfuscated with a small increase in area.

As opposed to existing logic locking techniques, measuring bit-level corruption at the outputs is not an appropriate metric for a filter. Instead, to explore the impact of decoys on the filter behavior, its zero-phase frequency response is computed based on the original coefficients and the constants selected randomly from the original coefficients and decoys. Fig.~\ref{fig:zpfr} shows the behavior of the filter \textit{Maskell07\_A108} when $p$ is 128 and a correct (blue) and 100 wrong (red) keys are applied. 

Observe from Fig.~\ref{fig:zpfr} that the decoys can alter the filter design characteristics, obfuscating the filter behavior. Note that decoys can be selected to change the filter behavior completely under each possible wrong key, considering also the hardware complexity and resiliency to the logic locking attacks.

	\section{Conclusions}
\label{sec:conclusions}

This paper presented a hardware obfuscation technique that prevents an adversary at an untrusted foundry from reverse engineering the constants of a TMCM block, a fundamental operation in the folded design of many IPs, such as ANNs and DSP circuits. The proposed technique obfuscates the target constants of the TMCM block using unique decoy constants and additional logic with keyed inputs. The obfuscation process takes place at RTL before logic synthesis, rather than at gate-level as in traditional logic locking schemes. It is observed that the proposed technique generates obfuscated TMCM designs that are resilient to well-known logic locking attacks. The proposed design architectures introduce alternative realizations of the TMCM block with different hardware complexity and resiliency to the attacks, enabling a designer to choose the one that fits best in an application. In any case, \textbf{the true IP} of these designs, i.e., the constants themselves, can be protected.

As a future work, the impact of synthesis optimizations on the hardware complexity and resiliency of the obfuscated designs is being explored.

\pagebreak
	\section*{Acknowledgment}

This work has been partially conducted in the project ``ICT programme'' which was supported by the European Union through the European Social Fund. It was also partially supported by European Union's Horizon 2020 research and innovation programme under grant agreement No 952252 (SAFEST) and by the Estonian Research Council grant MOBERC35.
	
	\bibliographystyle{IEEEtran}
	\bibliography{iolts21}
	
\end{document}